\documentstyle[prl,aps,epsf,preprint]{revtex}

\preprint{\vbox{\noindent July 1995\hfil astro-ph/9507023\break
Revised February 1996\hfil MPI-PTh/95-63\break
\hbox{\ }\hfil SFB-375/18\break
\hbox{\ }\hfil BA-95-24\break
\hbox{\ }}}

 \draft

 \tighten

\begin{document}
\title{Nucleon Spin Fluctuations and the Supernova Emission of
Neutrinos and Axions\footnote{To be published in Physical Review Letters}}
\author{Hans-Thomas Janka$^*$}
\address{Department of Astronomy and Astrophysics,
The University of Chicago, Chicago, Illinois 60637-1433}
\author{Wolfgang Keil}
\address{Max-Planck-Institut f\"ur Astrophysik,
Karl-Schwarzschild-Str.~1, 85740 Garching, Germany}
\author{Georg Raffelt}
\address{Max-Planck-Institut f\"ur Physik,
F\"ohringer Ring 6, 80805 M\"unchen, Germany}
\author{David Seckel}
\address{Bartol Research Institute,
University of Delaware, Newark, Delaware 19716}
\date{\today}
\maketitle
\begin{abstract}
In the hot and dense medium of a supernova (SN) core, the nucleon
spins fluctuate so fast that the axial-vector neutrino opacity and the
axion emissivity are expected to be significantly modified. Axions
with $m_a\alt10^{-2}\,{\rm eV}$ are not excluded by SN~1987A. A
substantial transfer of energy in neutrino-nucleon ($\nu N$)
collisions is enabled which may alter the spectra of SN neutrinos
relative to calculations where energy-conserving $\nu N$ collisions
had been assumed near the neutrinosphere.
\end{abstract}
\pacs{PACS numbers: 14.60.Lm, 14.80.Mz, 95.30.Cq, 97.60.Bw}

\begin{figure}
\centering\leavevmode
\epsfxsize=4in
\epsfbox{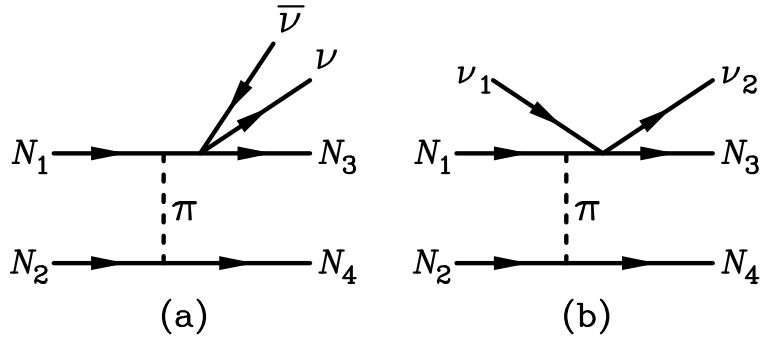}
\bigskip
\caption{(a)~Bremsstrahlung emission of neutrino pairs in $N\!N$
collisions.  Instead of a neutrino pair, an axion can be emitted.
(b)~``Inelastic $\nu N$ scattering'' in the presence of a bystander
nucleon.}
\label{F1}
\end{figure}

{\it Introduction.}
Nucleons interact with each other by a spin-dependent force which
causes their spins to flip in a typical nucleon-nucleon ($N\!N$)
collision. In the hot nuclear medium of a supernova (SN) core where
$\rho=10^{14}$--$10^{15}\,\rm g\,cm^{-3}$ and $T=10$--$60\,{\rm MeV}$
the spin-fluctuation rate $\Gamma_\sigma$ is estimated to be at least
of order $T$. Fluctuating spins act as a source for radiation such as
neutrino pairs or axions which couple to the nucleon spin. Also,
particles which scatter on nucleons by a spin-dependent force such as
neutrinos or other nucleons can gain or lose energy if the target spin
fluctuates. Further, fast spin fluctuations cause an averaging of the
local spin density which reduces the effective $\nu N$ or $N\!N$
scattering cross section. To lowest order the physical essence of
these processes is described by the Feynman graphs of 
Fig.~\ref{F1}.\footnote{Fig.~\ref{F1}b does not provide a complete
lowest-order description of ``inelastic $\nu N$ scattering'' because
one needs to include wavefunction and vertex renormalization terms
\cite{Sawyer95}. However, our use of a sum rule obviates the need to
include these terms explicitly \cite{RSS}.} In a series of papers
\cite{RSI,RSII,KJR} we have called attention to some of these
phenomena and their consequences for the SN emission of neutrinos and
axions. In the present note we provide a consistent account of this
research and summarize its implications.

{\it The Spin-Density Structure Function.}
We focus on an isotropic, nonrelativistic, nondegenerate medium of
baryon density $n_B$, temperature $T$, and a single species of
nucleons. The axial-vector neutrino interaction with this medium is
governed by the dynamical spin-density structure function (e.g.\ 
Refs.~\cite{RSII,IP})
\begin{equation}
S_\sigma(\omega,{\bf k})=
\frac{4}{3n_B}\int_{-\infty}^{+\infty} dt\,
e^{i\omega t}\langle\hbox{\boldmath$\sigma$}(t,{\bf k})\cdot
\hbox{\boldmath$\sigma$}(0,-{\bf k})\rangle
\label{E1}
\end{equation}
where $\hbox{\boldmath$\sigma$}(t,{\bf k})$ is the spatial Fourier
transform at time $t$ of the nucleon spin-density operator
$\hbox{\boldmath$\sigma$}(x)\equiv\frac{1}{2}\psi^\dagger(x)
\hbox{\boldmath$\tau$}\psi(x)$. Here $\psi(x)$ is the nucleon field
(a Pauli two-spinor) and $\hbox{\boldmath$\tau$}$ is a vector of Pauli
matrices. The expectation value $\langle\ldots\rangle$ is taken over a
thermal ensemble so that detailed balance
$S_\sigma(\omega,{\bf k})=S_\sigma(-\omega,-{\bf k})\,e^{\omega/T}$ is
satisfied.

The axial-current transition rate for a neutrino of four momentum
$(\varepsilon_1,{\bf k}_1)$ to $(\varepsilon_2,{\bf k}_2)$ is then
found to be $\frac{1}{4}C_A^2G_{\rm F}^2 n_B
(3-\cos\theta)\,S_\sigma(\omega,{\bf k})$ with $\theta$ the scattering
angle,
$(\omega,{\bf k})=(\varepsilon_1-\varepsilon_2,{\bf k}_1-{\bf k}_2)$,
and $G_{\rm F}$ the Fermi constant \cite{RSII}. The axial 
neutral-current coupling $C_A$ to free $p$'s and $n$'s should differ 
somewhat from the charged-current value 1.26 because of the 
polarization of strange quarks in nucleons. Further, a suppression by
$\approx 20\%$ of all axial couplings is likely in a bulk nuclear 
medium \cite{RSII}.

We further neglect the momentum transfer from neutrinos to
nonrelativistic nucleons (long-wavelength approximation) so that only
$S_\sigma(\omega)=\lim_{{\bf k}\to0}S_\sigma(\omega,{\bf k})$ is used
\cite{RSII}.  After an angular integration the axial-current
scattering cross section~is
\begin{equation}
\frac{d\sigma_A}{d\varepsilon_2}=\frac{3C_A^2G_{\rm F}^2}{4\pi}\,
\frac{\varepsilon_2^2\,S_\sigma(\varepsilon_1-\varepsilon_2)}{2\pi}.
\label{E2}
\end{equation}
This result is based on the interaction Hamiltonian
$(C_AG_{\rm F}/2\sqrt2)\,\overline\psi_N\gamma_\mu\gamma_5\psi_N\,
\overline\psi_\nu\gamma^\mu(1-\gamma_5)\psi_\nu$. Axions $a$ couple to
nucleons by
$(C_N/2f_a)\,\overline\psi_N\gamma_\mu\gamma_5\psi_N\,\partial^\mu a$
where $C_N$ is a model-dependent factor and $f_a$ the Peccei-Quinn
scale \cite{axions}. The axionic energy-loss rate per nucleon is
\cite{RSII}
\begin{equation}
\frac{Q_a}{n_B}=\frac{C_N^2}{(4 \pi)^2 f_a^2}
\int_0^\infty d\omega\,\omega^4\,S_\sigma(-\omega).
\label{E3}
\end{equation}
Within our approximations, $Q_a$ and $\sigma_A$ depend on the same
function $S_\sigma(\omega)$. For a mixture of $n$ and $p$ the nuclear
response is characterized by a linear combination of a small number of
functions to allow for the isospin dependence of different processes.

In a noninteracting medium the nucleon spins do not evolve, and so in
Eq.~(\ref{E1}) $\hbox{\boldmath$\sigma$} (t)=\hbox{\boldmath$\sigma$}
(0)$.  Then the time integration yields
$S_\sigma(\omega)=2\pi\,\delta(\omega)$.  Indeed, if one ignores
recoil effects the neutrino cannot gain or lose energy in a $\nu N$
collision. Also, $Q_a=0$ in agreement with the fact that free nucleons
do not emit radiation.

Nucleon-nucleon collisions with a spin-dependent force cause a
nontrivial evolution $\hbox{\boldmath$\sigma$}(t)$. Still, in
$\int_{-\infty}^{+\infty} d\omega\,S_\sigma(\omega)$ the
$e^{i\omega t}$ factor gives us $2\pi\delta(t)$. Therefore, one finds
that the ``sum''
$\int_{-\infty}^{+\infty}d\omega\,S_\sigma(\omega)=
(4/3 n_B)\,2\pi\langle\hbox{\boldmath$\sigma$} (0)\cdot
\hbox{\boldmath$\sigma$} (0)\rangle$ is independent of the
time evolution of $\hbox{\boldmath$\sigma$}(t)$.  Therefore,
unless the $N\!N$ interaction establishes spin-spin correlations which
are absent in a noninteracting medium, one finds the sum rule
$\int_{-\infty}^{+\infty}d\omega\,S_\sigma(\omega)=2\pi$.

In Ref.~\cite{Sawyer} the static spin-density structure function
$S_\sigma({\bf k})=\int_{-\infty}^{+\infty} S_\sigma(\omega,{\bf k})\,
d\omega/2\pi$, and notably its long-wave\-length limit ${\bf k}\to 0$,
has been used to study correlation effects for SN neutrino opacities.
We believe that it is crucial to keep the dynamical structure because
the $\omega$-dependence is not essentially $\delta(\omega)$. Still,
the results of \cite{Sawyer} indicate a significant reduction of
$S_\sigma({\bf k})$ due to $N\!N$ interactions so that the partial
pairing of spins may reduce
$\int_{-\infty}^{+\infty}d\omega\,S_\sigma(\omega)$ below its free
value $2\pi$.

{\it Ansatz for $S_\sigma(\omega)$.} 
Motivated by a classical brems\-strah\-lung calculation we assume a
functional form
$S_\sigma(\omega)=\Gamma_\sigma\omega^{-2}s(\omega/T)\,b(\omega/T)$
where $\Gamma_\sigma$ is the spin-fluctuation rate, $s(x)$ a
dimensionless even function normalized such that $s(0)=1$, and
$b(x)=1$ for $x>0$ and $e^x$ for $x<0$ to satisfy detailed balance. If
one models the $N\!N$ interaction by a one-pion exchange (OPE)
potential one finds for a single species of nucleons \cite{RSII,BT}
\begin{equation}
\Gamma_{\sigma, {\rm OPE}} = 4\sqrt\pi\,\alpha_\pi^2
\frac{n_B T^{1/2}}{m_N^{5/2}}
= 8.6\,{\rm MeV}\,\rho_{13}\,T_{10}^{1/2},
\label{E4A}
\end{equation}
where $\alpha_\pi\equiv(f 2m_N/m_\pi)^2/4\pi\approx15$ with $f\approx
1.0$ is the pion fine structure constant,
$\rho_{13}\equiv\rho/10^{13}\,\rm g\,cm^{-3}$,
$T_{10}\equiv T/10\,{\rm MeV}$, and the vacuum value
$m_N=940\,\rm MeV$ has been used. In the following we shall use the 
classical result $s(x)=1$. The detailed large-$x$ behavior depends
on the assumed short-range behavior of the $N\!N$ interaction
potential, but the OPE potential is too singular at short distances
to yield a useful answer.

\begin{figure}
\centering\leavevmode
\epsfxsize=4in
\epsfbox{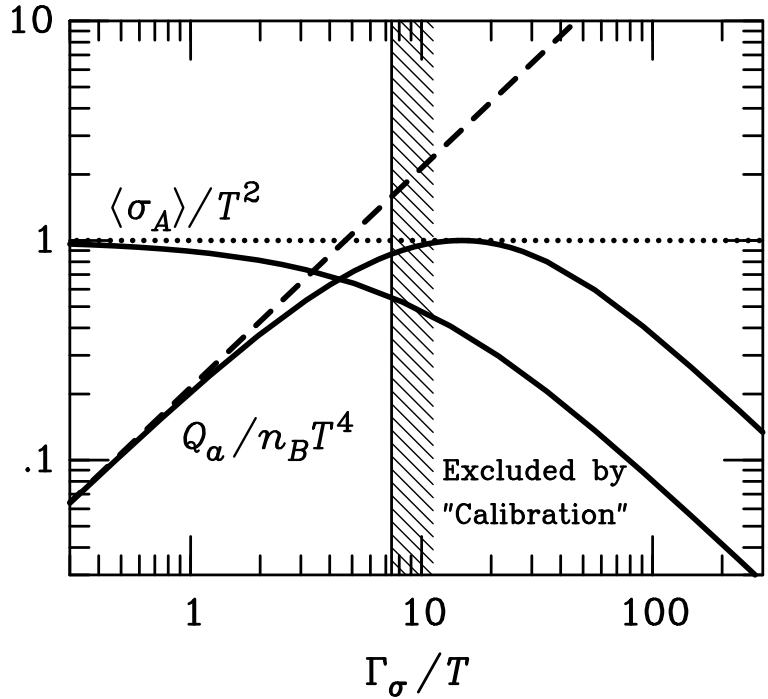}
\bigskip
\caption[...]{Schematic dependence of the average axial-vector $\nu N$
scattering cross section $\langle \sigma_A\rangle$ and of the axion
emission rate per nucleon $Q_a/n_B$ on $\Gamma_\sigma$. The broken
lines correspond to the naive results which are characterized by
$\Gamma=0$ in Eq.~(\ref{E4}) for $Q_a/n_B$ and by
$S_\sigma(\omega)=2\pi\delta(\omega)$ for $\langle\sigma_A\rangle$.
\label{F2}}
\end{figure}

For $\omega\alt\Gamma_\sigma$ the behavior of $S_\sigma(\omega)$ is 
dominated by multiple-scattering effects. A proper calculation
in this regime does not exist, but two of us (RS) have previously
argued that a plausible representation is \cite{RSI,RSII}
\begin{equation}
S_\sigma(\omega)\to\frac{\Gamma_\sigma}{\omega^2+\Gamma^2/4}\,
s(\omega/T), \label{E4}
\end{equation}
with $\Gamma \approx \Gamma_\sigma$. Taking $s(x) = 1$ the sum
rule is satisfied if one chooses $\Gamma=\Gamma_\sigma$ for
$\Gamma_\sigma\ll T$, and $\Gamma = \Gamma_\sigma/2$ for
$\Gamma_\sigma\gg T$. 
Assuming this form for $S_\sigma(\omega)$ allows us to compute the
variation of the thermally averaged axial-current scattering cross
section $\langle\sigma_A\rangle$, and of $Q_a/n_B$, with
$\Gamma_\sigma$ (Fig.~\ref{F2}). In a dilute medium
($\Gamma_\sigma\ll T$), $\langle\sigma_A\rangle$ is independent of
density. The axion emission rate per nucleon $Q_a/n_B$ is proportional
to $n_B$ and thus to $\Gamma_\sigma$. Such conditions may be relevant
near the ``neutrinosphere'' where $\rho_{13} \approx 0.1$. However,
for the conditions in a SN core Eq.~(\ref{E4A}) yields
$\Gamma_{\sigma,{\rm OPE}}/T=20$--$50$~\cite{KJR}.  While we will
argue that this perturbative result likely is an overestimate of the
true $\Gamma_\sigma$ it still indicates that multiple-scattering
effects may be expected to suppress both $\langle\sigma_A\rangle$ and
$Q_a/n_B$.

{\it ``Calibration'' of $\Gamma_\sigma$ at High Density.}
In the ansatz Eq.~(\ref{E4}) for $S_\sigma$ it was implicitly assumed
that $\Gamma_\sigma$ increases linearly with density, as for
$\Gamma_{\sigma,{\rm OPE}}$. It is unlikely that this is correct---a
similar ``averaging effect'' which reduces the $\nu N$ cross section
also affects $N\!N$ collisions. For $\Gamma_{\sigma, {\rm OPE}} \agt
T$, one should consider the true $\Gamma_\sigma$ to be a slowly
increasing, but unknown, function of the density.

The neutrino signal of SN~1987A offers a unique possibility to test
for the actual behavior of~$\Gamma_\sigma$. The neutrino signal of a
SN depends sensitively on the neutrino opacity at high density, which
is dominated by $\sigma_A$. Since $\sigma_A$ drops precipitously if
$\Gamma_\sigma\agt T$, we in principle have an experimental probe of
$\Gamma_\sigma(\rho,T)$. To this end, three of us (KJR) have performed
detailed numerical calculations of the cooling of newborn neutron
stars with modified axial-vector neutrino opacities \cite{KJR}.  The
result of this study is that reduced neutrino opacities shorten the
neutrino pulse dramatically and increase the number of events and the
event energies in the Kamiokande~II and IMB detectors. The best fit to
the data is achieved if one ignores opacity suppression effects
entirely. An overall reduction by more than 50\% seems to be excluded.
This is still compatible with the expected reduction of $C_A$ in a
bulk nuclear medium which entails an opacity suppression by 30--40\%
\cite{RSII}, but would be incompatible with any significant reduction
due to either ``pairing'' of nucleon spins by the $N\!N$ interaction
or broadening of $S_\sigma(\omega)$. Similar conclusions follow from
Ref.~\cite{BurrowsRT} where the opacity was indirectly lowered by
increasing the number of neutrino~species.

Of course, the SN~1987A observations encompass very few events.
Probably the most reliable measure of the opacities in the deep
interior is the neutrino pulse duration as the number of events and
their energies are largely determined by the spectral features of the
neutrino flux and thus by the conditions near the neutron star
surface. It is worrying that a long signal and thus large opacities
are mostly supported by the three late Kamiokande events which are
separated from the first eight by a $7\,\rm s$ break. If the late
events had some other cause besides the Kelvin-Helmholtz cooling such
as late-time accretion, possibly triggered by a reverse-shock
\cite{JankaSFB}, then reduced opacities might become more tolerable.

Barring this possibility we infer that $\Gamma_\sigma$ cannot grow
linearly with density but reaches a value of at most a few times $T$
in a SN core. Values for $\Gamma_\sigma$ to the right of the hatched
band in Fig.~\ref{F2} seem to be excluded. If correct, this result
implies that the OPE value for $\Gamma_\sigma$ is an overestimate.

{\it Axion Bounds.} 
Previous axion bounds from SN~1987A had been based on the ``naive
emission rate'' (dashed line in Fig.~\ref{F2}), and had employed a
perturbative estimate of $\Gamma_\sigma$. During the first few seconds
after collapse $\Gamma_{\sigma,{\rm OPE}}/T=20$--50 throughout the SN
core. This would take one somewhat to the right of the maximum of the
solid line in Fig.~\ref{F2}. If our ``calibration'' of $\Gamma_\sigma$
is correct, one is somewhat to the left of the maximum. Either way,
$Q_a$ is near its theoretical maximum so that the ``naive''~$Q_a$ is
suppressed by about a factor of 10. Because $Q_a$ is proportional to
the square of the axion mass, previous limits on $m_a$ are relaxed by
about a factor of~3.

The often-quoted limit $m_a\alt 10^{-3}\,\rm eV$ was based on
$C_N=0.5$ for $N=p$ and $n$ \cite{BTB}. In popular axion models these
couplings are much smaller, for example $C_n\approx0$ in the so-called
KSVZ model. Therefore, in a detailed review \cite{axions} it was shown
that translating the SN~1987A bound on the $aN$ coupling into a bound
on $m_a$ really yields $m_a\alt 3{\times}10^{-3}\,\rm eV$ in the KSVZ
and other typical models. Together with the above suppression effect
we conclude that $m_a\alt 10^{-2}\,\rm eV$ is not excluded.

{\it Energy Transfer in $\nu N$ Collisions.}
In previous SN studies of neutrino transport, $\nu N$ collisions have
been taken to be energy conserving \cite{Spectra,Janka} because
$S(\omega) \propto \delta(\omega)$ for scattering of $\nu$'s from
single, ``heavy'' nucleons. The $\nu_\mu$ (and $\overline\nu_\mu$,
$\nu_\tau$, and $\overline\nu_\tau$) spectra are formed at the radius
where $\nu e$ scattering and $\nu_\mu \overline\nu_\mu \leftrightarrow
e^-e^+$ become inefficient for transferring energy between the
neutrinos and the medium.  However, the $\nu_\mu$'s must diffuse
through a considerable overburden of nuclear material before they
escape. In general this material will have a temperature $T$ that is
lower than the temperature $T_\nu$ that characterizes the $\nu_\mu$
spectrum.  Thus the average $\nu_\mu$ energies are usually found to be
a factor of 1.3--1.8 larger than those of $\overline\nu_e$'s, which
react via $\beta$-processes and decouple at lower density and
temperature.

A broad $S_\sigma(\omega)$ such as in Eq.~(5) would allow for energy
transfer between the neutrinos and the medium. Further, even for $\nu
N$ processes the $\delta$ function is only an approximation that
neglects recoil corrections. Their inclusion would modify the expected
$\nu_\mu$, $\overline\nu_\mu$, $\nu_\tau$, and $\overline\nu_\tau$
spectra. In turn, energy conservation and a reduced energy flux in
these flavors would imply increased $\nu_e$ and $\overline\nu_e$
energy and number fluxes.  This is important in planning for detection
of SN neutrinos at Superkamiokande or SNO, for understanding the
explosion mechanism of Type~II supernovae \cite{BeWiJa,Fuller}, and
for the conjecture that r-process nucleosynthesis occurs in
the high-entropy medium that surrounds the protoneutron star
\cite{Rproc}.

For neutrinos with a black-body spectrum the average recoil energy
transfer in $\nu N$ collisions with nondegenerate nucleons is
$\langle\Delta\varepsilon\rangle_{\nu N}\approx30\,T_\nu(T-T_\nu)/m_N$
\cite{Janka,Tubbs}, and the average energy exchange
$\langle\Delta \varepsilon\rangle_{\nu N\!N}$ in inelastic $\nu N\!N$
scatterings is
\begin{equation}
\frac{\int_0^\infty d\varepsilon_1\varepsilon_1^2
e^{-\varepsilon_1/T_\nu} \int_0^\infty
d\varepsilon_2\varepsilon_2^2(\varepsilon_2-\varepsilon_1)
S_\sigma(\varepsilon_1-\varepsilon_2)}
{\int_0^\infty d\varepsilon_1\varepsilon_1^2 e^{-\varepsilon_1/T_\nu}
\int_0^\infty d\varepsilon_2\varepsilon_2^2
S_\sigma(\varepsilon_1-\varepsilon_2)}.
\label{E6}
\end{equation}
With Eq.~(\ref{E4}), the sum rule for $S_\sigma(\omega)$, and $s(x)=1$
we find numerically $\langle\Delta\varepsilon\rangle_{\nu N\!N}=
\zeta\,\frac{1}{4}\Gamma_\sigma\,(T-T_\nu)(T_\nu T)^{-1/2}$ where
$\zeta^{-1}\approx1+\log_{10}(1+\Gamma_\sigma/T)$ within 15\% for
$1<T_\nu/T<2$ and $0.1<\Gamma_\sigma/T<10$. For conditions near the
neutrinosphere both the $\nu N$ and $\nu N\!N$ processes can therefore
transfer several MeV of energy to the medium.  Even though the energy
exchange per interaction may be small, a typical neutrino experiences
10--100 interactions during the diffusion process through the
protoneutron star surface layers. The net effect may lead to
substantially lower average energies of the $\nu_\mu$'s and
$\overline\nu_\mu$'s. Thus, both $\nu N$ recoil and the inelastic 
$\nu N\!N$ modes of energy transfer are potentially important near the
neutrinosphere where the spectra are formed.

The same conclusion is reached by comparing the recoil energy transfer
rates between neutrinos and a medium of nucleons and between neutrinos
and a medium of degenerate electrons \cite{Janka},
\begin{equation}
\frac{Q_{\nu N}}{Q_{\nu e}}\approx
0.8\,\frac{1 + 5C_A^2}{2 C_e}\,\rho_{13}^{-1/3}
\left(\frac{T_\nu}{10\,{\rm MeV}}\right)^{\! 2}\!
\left(\frac{Y_e}{0.1}\right)^{\! -4/3} \,.
\label{E8}
\end{equation}
Here, the effective weak interaction constant $C_e$ for $\nu e$
scattering is 0.33 ($\nu_{\mu,\tau}$), and 0.26
($\overline\nu_{\mu,\tau}$). In the neutron star surface layers $\rho
\approx 10^{12}$--$10^{13}\,{\rm g/cm}^3$, and $Y_e$ drops to
0.05--0.1 within about a second after stellar core collapse
\cite{BuWiKe}.  The energy transfer from $\nu_\mu$ to the medium is
apparently dominated by direct transfer to the nucleons by an order of
magnitude or greater even though the calculation behind Eq.~(\ref{E8})
does not include the full effects of electron phase-space blocking.

{\it Summary.}
For axial-vector current interactions of neutrinos and axions in a SN
core multiple-scattering effects must not be ignored. Notably, the
spin-fluctuation rate cannot be simultaneously very small
($\Gamma_\sigma\ll T$), as had been assumed for the neutrino
opacities, and very large ($\Gamma_\sigma\gg T$), as had been assumed
for the axion emissivities. Our study suggests that it is in the
middle, i.e.\ of order $T$, a range where the impact on neutrino
opacities is relatively mild while the axion emissivity is near its
maximum allowed by multiple-scattering effects (but still less than
previous calculations had indicated). At the present state of the art,
the neutrino opacities cannot be calculated from first principles
within controlled approximations. Therefore, the SN neutrino signal
duration and other signal characteristics can be predicted only with
the accuracy of a dimensional analysis, implying uncertainties for
models of the $\nu$-driven mechanism of Type~II SN explosions
\cite{BeWiJa}. Even near the neutrinosphere nonnegligible effects such
as the nonconservation of energy in $\nu N$ collisions are important
where both the recoil and the bremsstrahlung mode of $\nu N$ energy
transfer need to be included.  A quantitative understanding of the
impact of energy transfer in $\nu N$ collisions on the expected
neutrino fluxes and spectra will require detailed numerical
simulations.

{\it Note Added.} 
After this paper had been submitted, a perturbative calculation of the
reduction of weak scattering rates by nucleon (iso)spin fluctuation
has appeared \cite{Sawyer95}.  In essence, this is a calculation of
the slope of our $\langle \sigma_A\rangle$ curve in Fig.~\ref{F2} at
the point $\Gamma_\sigma/T=0$.  In linear-response theory, this slope
is independent of our Lorentzian ansatz and agrees with the
perturbative result---for a detailed comparison see Ref.~\cite{RSS}.
Further, Sigl \cite{Sigl95} has derived an f-sum rule for
$S_\sigma(\omega)$ which allows one to estimate an upper limit on
$\Gamma_\sigma$, providing a theoretical underpinning for our SN~1987A
bound on $\Gamma_\sigma/T$.

{\it Acknowledgements.}
At the University of Chicago this research was supported by
NSF grant AST 92-17969, NASA grant NAG 5-2081, and by an Otto Hahn
Postdoctoral Scholarship of the Max-Planck-Society. At the
Max-Planck-Institutes, partial support by the European Union contract
CHRX-CT93-0120 and by the Deutsche Forschungsgemeinschaft grant SFB
375 is acknowledged, and at Bartol by DOE grant DE-AC02-78ER05007.



\end{document}